\documentclass{article}
\begin{document}
\vbadness = 100000
\hbadness = 100000
\title{The invariance of the speed of light}
\author{Jerrold Franklin\footnote{Internet address:
Jerry.F@TEMPLE.EDU}\\
Department of Physics\\
Temple University, Philadelphia, PA 19122-6082}
\maketitle

\begin{abstract}
We show that the conclusion of a recent experiment\cite{g} that claims to have discovered that
``the speed of light seems to depend on the motion of the observer" is wrong.  
\end{abstract}

A recent paper\cite{g} claims to have measured an increase in the speed of light\cite{g2}
due to motion of the detector of the light.  The theory of special relativity is based on the postulate that the speed of light is a constant, which cannot depend on the motion of the detector.  This means that a positive effect of the motion of the detector on the measured speed of light, as claimed in \cite{g}, would refute special relativity and much of modern physics.  Fortunately for relativity, the time of flight measurement made in \cite{g} has been misinterpreted in that paper.  We show in this paper that the measurements in \cite{g}  do not, in fact, measure any effect of the motion of the detector.  Any indication of a change in the speed of light claimed in \cite{g} has entered through the faulty theoretical analysis in \cite{g} of the measurements made there.

The first thing to note is that the only measurement actually made in \cite{g} was of the elapsed time between the emission of a laser pulse from an emitter at rest in an observatory on the surface of the Earth in a Lorentz frame O, and its absorption in a detector at rest in the same observatory.  The laser pulse was reflected back to the observatory by a reflector on the Moon, which was moving toward the emitter and detector in frame O with a speed $v_0$ due to the rotation of the Earth.  This single time of flight measurement was then analyzed in two separate Lorentz frames, O, and a frame S in which the center of the Earth and the center of the Moon were each at rest, but in which the emitter and detector were moving with speed $v_0$.  That is, the same time of flight measurement was used to get two different values for the speed of light.  The value found in frame S  was in approximate agreement with the standard value for $c$.
But the speed of light calculated by \cite{g} in frame O was about 200 m/s higher. 
Since 200 m/s was close to the velocity of the detector toward the Moon at the time of measurement,
\cite{g} attributed this difference to the motion of the detector.
 We show below that it was a faulty calculation in the second case that led to the incorrect value for the speed of light.

  For one thing, in order to measure a difference in the speed of light measured by a moving detector compared to a detector at rest, two separate speed measurements should be made on two detectors, one at rest and one moving in the same Lorentz frame.  You can only measure a difference if there are two different measurements to compare.  Yet \cite{g} claims to have measured this difference with only one measurement, analyzed in two different ways.  That is, it was a difference in the two calculations that was found by \cite{g}, not a difference in two different measurements.

In Sec. 4.2 of \cite{g}, the velocity
 $c_{\rm O}$ is calculated as the presumed speed of light measured in the frame O in which both the emitter and detector are at rest.  Although the calculation is done in a frame in which the detector is at rest,  \cite{g} calls this a measurement of the speed of light by a moving detector.  This is because the observatory is moving with respect to the centers of the Earth and the Moon, whose mutual rest system (neglecting their relative motion) is designated by \cite{g} as ``the local stationary frame (S)".  Presumably, the emitter and detector are in some form of absolute motion with respect to this ``local stationary frame", and presumably due to this motion the speed of light in frame O is an amount $v_0$ greater than the speed of light calculated in frame S. 

Where is the error in the calculation in \cite{g} of the speed of light in frame O?  
First we present a simple derivation that gives the correctly measured speed of light in frame O. 
In frame O, the rest frame of the emitter and the detector, the reflector on the Moon is moving toward the emitter and detector with a speed $v_0$ that corresponds to the speed of the observatory on the surface of the Earth due to the Earth's rotation.  (For simplicity we consider the velocity of the reflector to be directed directly toward the emitter, although vector addition would be required in general.)  
With no motion of emitter, detector, or reflector, the speed of light calculated from a measured time of flight $T$ would be given by
\begin{equation}
c=2D/T,
\end{equation}
where $D$ is the distance of the reflector from the observatory.

When calculated in frame O, where the reflector on the Moon is moving toward the stationary emitter and detector, the reflector will have moved a distance $\Delta D/2=v_0 T/2$ toward the emitter when the laser pulse reaches it.  Thus the round trip distance followed by the laser pulse (or any other wave packet or material object) will be diminished by an amount $\Delta D$.  This decrease in distance should be included in calculating the speed of the pulse.  There will also be a decrease $\Delta T$ in the measured time of flight, given by
$\Delta T=\Delta D/c$  (neglecting here any slight change in the speed of light).
Thus the speed of light calculated in frame O should be
\begin{equation}
 c_{\rm O}=\frac{2D-\Delta D}{T-\Delta T}
=\frac{2D-\Delta D}{2D/c-\Delta D/c}=c, 
\end{equation}
This equation is the same as the equation used in \cite{g} to correctly calculate $c_{\rm S}$, the speed of light in frame S, with $\Delta D$ interpreted there as the distance moved by the detector in that frame.  The equality of this result for $c_{\rm O}$ with that for $c_{\rm S}$ in \cite{g} means that the same procedure to calculate the speed of light from the time of flight measurement should be used in each reference frame, precluding any difference in the measured speed of light.  

Reference \cite{g} refutes this simple derivation of $c_{\rm O}$ with the statement: ``In fact, the length of the outbound leg from launch to bounce actually increases due to motion of the retro-reflector toward the observer during the observation".  I present below a direct quote of the argument given for this, since I do not know how to put it in my own words:
``The retro-reflector folds the optical path (and frame O) back on itself. Motion of the retroreflector
toward the observer 'pushes' the folded segment of frame O continuously back,
behind the observer. By the time the bounce occurs the distance from the origin of the
pulse in frame O to the retro-reflector has increased from $D$ to $D+\Delta D$.  But this motion
simultaneously shortens the return leg by same the amount that it lengths the outbound
leg, so that when the bounce occurs the length of the return leg has decreased to
$D-\Delta D$ in frame O. ... Thus, even with the retro-reflector
moving in frame O during the measurement, the sum of the outbound and return legs in
frame O is the full initial distance $2D$."

As I interpret this argument, \cite{g} seems to be describing the motion of some virtual ``optical path" back on itself that has nothing to do with the real motion of the actual reflector moving toward the actual fixed emitter and detector.   In any event, \cite{g} describes an optical path that is not followed by the laser pulse.  It seems clear that if the actual reflector moves distance $\Delta D/2$ towards the emitter, which does not move in its rest frame, then it is a distance $\Delta D/2$ closer to the emitter when the laser pulse strikes it.  No discussion of an optical path length, especially one not followed by the laser pulse, can change that. 
 Reference \cite{g} calls its conclusion ``completely counter-intuitive", but more importantly, it is completely wrong.  Using this argument, \cite{g} used $2D$ instead of 
$2D-\Delta D$ for the distance traveled by the laser pulse, leading to its Eq. (3) for the `measured' speed of light
\begin{equation}
 c_{\rm O}=\frac{2D}{T-\Delta T}=c+v_0
\end{equation}
instead  of the correct equation (2).  Consequently, \cite{g} gets a value for $c_{\rm O}$ that is too large  by the speed $v_0$.
Note that the discrepancy, $v_0$, between the speed of light as calculated by \cite{g} in the two frames is independent of the measured time of flight $T-\Delta T$.

In conclusion, because of the error we point out above, the claim in \cite{g} that the velocity of light as measured by a moving detector does not equal $c$ is wrong.  It is particularly striking that the so called `measured' discrepancy of 200 m/s between the light speed as calculated in the two frames
does not depend at all on the measured time of flight of the laser pulse.  That is, using the procedure of \cite{g} to find $c_{\rm O}$, any random number put in for the time of flight $T$ would give a result that is $v_0$ larger than $c_{\rm S}$.  Since the claimed anomalous result does not depend on the recorded time of flight, this experiment has, in fact, measured nothing.


\begin{thebibliography}{9}
\bibitem{g2}Although we feel it is necessary to correct the faulty analysis in \cite{g}, the measurements made by \cite{g} could not have measured the speed of light even if analyzed correctly.  That is because all the accurate distances used in the analysis were, in fact, determined by the same type of laser pulse time of flight measurement used in \cite{g}.  The distances were all determined using the value  $c=299,792,458 m/s$ for the speed of light.
That means that, at best, all \cite{g} could be measuring was the combined accuracy of the earlier laser time of flight measurements and those in \cite{g}.
\end{thebibliography}
\end{document}